\documentstyle[12pt]{article}
\input{epsf}

\def\be{\begin{equation}}
\def\ee{\end{equation}}
\def\ba{\begin{eqnarray}}
\def\ea{\end{eqnarray}}

\def\12{{1\over 2}}

\def\msun{M_\odot}

\def\eg{{\it e.g.~}}
\def\ie{{\it i.e.~}}
\def\etal{{\it et~al.~}}
\def\ltsima{$\; \buildrel < \over \sim \;$}
\def\simlt{\lower.5ex\hbox{\ltsima}}
\def\gtsima{$\; \buildrel > \over \sim \;$}
\def\simgt{\lower.5ex\hbox{\gtsima}}

\begin{document}


\title{\bf Weakly mass-loaded accretion disks}
\author{Yuri A. Shchekinov\thanks{yus@phys.rnd.runnet.ru}\\
Department of Physics, University of Rostov,\\
Rostov on Don, 344090 Russia}
\maketitle

\begin{abstract}

Accretion disks with an additional mass input on the disk surface 
from environment are considered in the limit of low mass input rate, \ie 
when the accretion flow remains keplerian. Due to dissipation of kinetic 
energy of the infalling gas disk temperature increases, and can deviate 
significantly from standard temperature of reprocessing accretion 
disks in their outer regions. This increase in temperature produces an 
excess of emission in a long-wavelength range of the disk spectrum. An 
illustrative example of the spectrum of a weakly mass-loaded disk (the 
surface mass input rate $\dot M_{\rm load}\simeq 1.5\times 10^{-8}~\msun$ 
yr$^{-1}$) reasonably reproducing FIR excess observed in CS Chameleon is 
given. 

\end{abstract}
\newpage


\section{Introduction}
\noindent

Dynamics of the multi-phase interstellar medium is dominated by mass and 
momentum exchange between different phases. Ejection of mass into diffuse 
component from a condensed phase changes velocity of the bulk flow, 
increases pressure in diffuse phase, and heats diffuse gas due to 
conversion of kinetic energy of evaporated/destroied clouds into thermal 
energy of bulk medium (Cowie, McKee \& Ostriker 1981, Hartquist \etal 
1986). As a result, loaded flows (\ie the flows dominated by mass 
input from a condensed phase) can differ qualitatively from the ones in 
a homogeneous (monophase) medium. In a simplest approach mass loaded flows 
are described by dynamical equations with source terms. For accretion disks 
mass loading corresponds physically to a continuing mass infall on the disk 
surface from the residual gas envelope. 

Extremely inhomogeneous structure of 
the diffuse interstellar gas and molecular clouds shows self-similarity on the 
scales from 0.02 (the resolution threshold) to 100 pc 
(see, Falgarone, Phillips \& Walker 1991). It seems obvious 
that such self-similarity can extend to much smaller scales 
corresponding to protoplanetary condensations in accretion 
disks. [Note that the resolution limit, 0.02 pc, corresponds 
to planetary masses of clumps, $\sim 10^{-4}~\msun$]. One can expect 
that in such conditions clumpy gas from debries of the parent molecular 
cloud will provide a continuing matter infall on the accretion protoplanetary 
disk being already formed. In theoretical treatment of protoplanetary 
disks this circumstance is described by the source terms in mass and 
momentum equations (see, \eg Safronov \& Vityazev 1983, Papaloizou \& 
Lin 1995). Multiple CI and CO velocity components (Roberge \etal 2000), CaII 
absorptions (Barnes, Tobin \& Pollard 2000) and emission in FeII, SI, SiII, 
NiII and CI lines (Lecavelier des Etangs \etal 2000), observed in 
$\beta$ Pictoris, are seemingly connected with the infall of clumpy gas 
on the disk. Similar detections of the infalling circumstellar gas have 
been reported for Herbig Ae/Be stars (Grinin, Natta \& Tambovtseva 1996). 
One can assume that this phenomenon represents late stages of matter infall 
from a highly inhomogeneous primordial circumdisk cloud onto the accretion 
disk. If this is the case, and protoplanetary disks are loaded by infalling 
and passing through them dense clumps, both vertical and radial 
structure of such disks, the accretion rate and the temperature 
distribution can differ considerably from those in homegeneous (monophase) 
disks. In this paper we analyze properties of steady disks, 
concentrating mainly on radial temperature distribution, in a simple 
model of weak mass loading, \ie when mass loading does not violate the 
keplerian flow of the bulk motion. In Section 2.1 we describe the model, 
in Section 2.2 we formulate basic equations describing radial structure 
of thin accretion disks for weak mass loading limit, in Section 2.3 we 
describe the solution of energy equation and find radial temperature 
distribution, and discuss possible observational manifestations, the 
effects of angular momentum input associated with mass loading on temperature 
distribution are discussed in Sec. 2.4, Section 3 summarizes the results.  


\section{Model and governing equations}
\noindent
\subsection{Model of weakly loaded disks}
\noindent

As a model of a mass-loaded disk which admits simple analytical description  
we will accept the following: an accretion disk is surrounded by a 
spherical ``subsystem'' of dense condensations formed of the debries of 
a parent cloud and orbiting 
around the central star (Fig. 1). This subsystem can be in a quasistationary 
state either because the condensations are nearly 
dissipationless, or an infinite 
(in practice, very massive) reservoir of gas replenishes dissipation of mass 
and energy. In the first case, the characteristic life time of 
the quasistationary orbits determined by interaction of the condensations 
with the diffuse component, with other condensations and with the disk itself, 
must be much larger than characteristic rotation period of the disk. For this 
to fulfill the following relations must hold, respectively 

\be
N_i\ll N_c, \quad \langle N_c\rangle \ll N_c, \quad N_d\ll N_c, 
\ee
where $N_i$, $N_c$, and $N_d$ are the column densities of the diffuse 
intercloud gas over the whole circumdisk cloud, of a single clump, 
and of the disk, respectively, 
$\langle N_c\rangle =fN_c$, $f={\cal N}_cR_c^2/R^2$ is the covering factor 
of clumps, 
$R_c$ is the clump radius, $R$, the radius of the circumdisk cloud. These 
inequalities 
definitely fulfill for planetesimals or comet-like bodies. For 
gaseous clumps it is 
not so obvious, and one has to assume a large enough density contrast 
in order that 
the clumps were long-living and orbiting quasistationary. 
In the latter case when the clumps are not sufficiently dense, so that 
for instance $N_c\sim N_d$ and 
they are absorbed by the disk in a single encounter, the overall 
picture can be kept 
quasistationary due to replenishment of the clumps from the 
surrounding molecular 
cloud as shown schematically on Fig. 1. A gas reservoir with radius 
$R\sim 3-10$ pc can support continuous mass infall during the time 
$t\sim 1-3$ Myr with the rate $10^{-7}-10^{-6}~\msun$ yr$^{-1}$ if the 
mean density is only $n_cf_v\sim 0.03-0.3$ cm$^{-3}$, where $n_c$ is gas 
density in clumps, $f_v$, the volume filling factor of the clumps; 
optical depth corresponding to the whole ensemble of such clumps is as 
small as $\tau_c\sim 10^{-21}n_cf_vR\sim 0.03$. As we will see below, mass 
loading even with one order of magnitude lower mass rate can produce 
observable effects. We will assume that the quasistationarity 
takes place in either way. With these assumptions the subsystem 
of clumps can be treated as stationary. When the clumps 
intersect the disk they lose a fraction of their mass due to stripping 
under the action of Rayleigh-Taylor instability (see detailed discussion 
in Hartquist \etal 1986, Klein, McKee \& Colella 1994). 
The subsystem of clumps is assumed in rest as a whole, \ie 
$\langle{\bf v_c(r)}\rangle =0$, where the averaging is over the 
ensemble of clumps at 
given $\bf r$, such that the angular momentum input into the disk 
from destroying clumps is zero. More general case with 
${\bf v}_c={\bf v}_c({\bf r})$ will be discussed in Sec. 2.4. 

\medskip
\noindent
\subsection{Equations of motions}
\noindent

In this framework the continuity eqation for thin disk is described 
by the equation 

\be
\label{mass}
{1\over R}{d\over dR}(R\Sigma u_R)=\Psi,
\ee
where $\Sigma=\int_{-\infty}^{+\infty}\rho dz$, 
$u_R=\int_{-\infty}^{+\infty}v_r dz$, 
(the interrelation $\Sigma u_R=\int_{-\infty}^{+\infty}\rho v_r dz$ is assumed 
as well), $\Psi=\int_{-\infty}^{+\infty}q dz$, 
$q$ [g cm$^{-3}$ s$^{-1}$] is the mass ejection rate from destroying 
clumps per unit volume. [In general, $\Psi$ may represent not only mass 
loading due to destroying cloudletts, but also mass infall onto the disk from 
the envelope (Safronov \& Vityazev 1983), the effects of the mass transfer 
stream in cataclismic variable stars, and the mass loss through a wind 
(Papaloizou \& Lin 1995)].  

Naviet-Stokes equation for azimuthal component of a stationary disk with 
zero angular momentum input is written in a standard form (see, Shakura \& 
Sunyaev 1973, Pringle 1981)  

\be
\label{vazimut}
{d\over dR}(\Sigma R^3\Omega u_R)={d\over dR}\left(
\nu\Sigma R^3{d\over dR}\Omega\right),
\ee
here $u_\phi=\Omega R$ is explicitly taken into account, $\Omega=\Omega(R)$ 
is the local angular velocity,  $\nu$ is the kinematical viscosity. 
The change of the angular velocity $\Omega$ in one rotational period 
$T_R=2\pi/\Omega$ due 
to mass input is $|\Delta \Omega|\sim \Psi/\Sigma$. From requirement 
that $|\Delta \Omega|/\Omega\ll 1$ one obtains restriction on the mass 
loading $\Psi\ll \Sigma\Omega/2\pi$, so that the ratio of the mass input rate 
$\dot M_p$ associated with the infalling clumps to the standard mass accretion 
rate $\dot M_0\sim 2\pi \nu\Sigma$ is 

\be
\label{criter}
{\dot M_p\over \dot M_0}\ll {1\over 4\pi}{R^2\Omega\over \nu}\sim 
{1\over 4\pi}{u_\phi\over |u_R|}\gg 1.
\ee
Obviously, the restrictions on the muss input $\Psi$ are rather weak. 
Formally, the change in $\Omega$ is larger on a much longer 
accretion time, 
$T_D=R/|u_R|\sim R^2/\nu$ (here $u_R\sim \nu/R$ is explicitly assumed, 
see Pringle 1981) 

\be
|\Delta\Omega|\sim \Omega_0 {\dot M_p\over \dot M_0},
\ee
where $\Omega_0$ is characteristic angular velocity in the absence of 
mass loading. 

For the radial velocity component we have  

\be
\label{vradial}
u_R{du_R\over dR}-{u_\phi^2\over R}+{1\over \rho}{dp\over dR}+
{GM\over R^2}=-{\Psi u_R\over \Sigma}.
\ee
Assuming that for a weak loading the radial component $u_R$ 
is of the order $\nu/R$ (see, Pringle 1981) we arrive at 

\be
\label{velofi}
u_\phi^2={GM\over R}\Bigl[1+O({\cal M}^{-2})+\beta(R)\Bigr],
\ee
where $\beta=\Psi u_RR^2/GM\Sigma$, 
${\cal M}=u_\phi/c_s$ is the local azimuthal Mach number, $c_s$, the local 
sound speed. Equations (\ref{mass})-(\ref{vradial}) differ from the 
standard equations 
of thin accretion disk (Shakura \& Sunyaev 1973, Lyndel-Bell \& Pringle 1974, 
see also reviews by Pringle 1981, and Papaloizou \& Lin 1995) by the sources 
due to mass loading in the r.h.s. of continuity equation (\ref{mass}) and in 
equation for radial velocity (\ref{vradial}). The parameter $\beta$ can be 
estimated as

\be
\label{acrpar}
\beta\sim {\alpha\over \pi}{T_R\over T_D}{\dot M_p\over \dot M_0}\ll 1, 
\ee
with the assumption that $\dot M_p/\dot M_0$ can have low or moderately 
high values, here $\alpha$ is the 
viscosity parameter (Shakura \& Sunyaev 1973), and thus its contribution 
to the r.h.s. of (\ref{velofi}) is negligible, and in this sense the loading 
will be treated as weak. It is readily seen that approximately 
$\beta\sim 2\pi \Psi/\Omega\Sigma$, and (\ref{acrpar}) is equivalent to 
the condition that on one rotational period $|\Delta \Omega|\ll \Omega$.
This means that the infalling settles onto the generic keplerian motion 
quickly, and as long as the mass input from clumpuscules is small 
enough on one rotational period, \ie (\ref{criter}) fulfills, it can be 
treated as a weak loading.

Defining the accretion rate as $\dot M=-2\pi R\Sigma u_R$ (see Pringle 1981) 
one can obtain the following solutions of equations (\ref{mass}) and 
(\ref{vazimut}) for a weakly loaded (\ie $\beta\ll 1$) disk 

\be
\dot M=\dot M_\ast-2\pi\int\limits_{R_\ast}^R \Psi R dR,
\ee

\be
\nu\Sigma={\dot M_\ast\over 3\pi}\Bigl[1-\left({R_\ast\over R}\right)^{1/2}
\Bigr]-{2\over 3}\int\limits_{R_\ast}^R \Psi RdR,
\ee
where $\dot M_\ast=-2\pi (R\Sigma u_R)_\ast$ is the total mass accretion rate 
(\ie the rate on the stellar surface); note, that the mass rate 
provided by the accretion disk itself in the absence of mass loading is equal 

\be
\dot M_0=\dot M_\ast-2\pi\int\limits_{R_\ast}^\infty \Psi RdR.
\ee
The quantity $\nu\Sigma$ determines dissipation at a rate 

\be
\label{disener}
D(R)=\12 \nu \Sigma(R\Omega')^2={3GM\dot M_\ast\over 8\pi R^3}
[1-(R_\ast/R)^{1/2}]-{3GM\over 4R^3}\int\limits_{R_\ast}^R\Psi(R)RdR,
\ee
and the corresponding luminosity integrated over the disk 

\be
L_D^\nu=4\pi\int\limits_{R_\ast}^\infty D(R) RdR=\12 
{GM\dot M_\ast\over R_\ast} -
3\pi GM\int\limits_{R_\ast}^\infty{dR\over R^2}\int\limits_{R_\ast}^R 
\Psi(R')R'dR'.
\ee

\subsection{Energy equation}
\noindent

One of the most interesting and important consequence of the mass loading  
is connected with possible heating of the disk 
due to conversion of kinetic energy of the falling gas into heat. The energy equation in the presence of mass loading
(see, Cowie, McKee \& Ostriker 1981, White \& Long 1991), integrated over $z$, 
is written as 

\be
\label{clumheat}
\Sigma\dot\varepsilon={3\over 2}{{\cal R}\over \mu}T\Psi\Bigl[
\12(\gamma-1){\rho\langle u^2\rangle\over p}-\gamma\Bigr]+\12 \nu\Sigma
(R\Omega')^2-{\cal L}(T,\rho),
\ee
where $\varepsilon$ is specific (per unit mass) thermal energy, the first 
term in the r.h.s. describes the energy input due to inelastic interaction 
of gas stripped from clumps and involved in keplerian motion of the disk, 
$\langle u^2\rangle$ is the mean square relative velocity of clumps and 
disk (the averaging is over the ensemble of clumps), ${\cal R}$ is gas 
constant, ${\cal L}$ is the 
radiative cooling rate, we will assume ${\cal L}=\sigma T^4$, 
$\sigma$ is Stefan-Bolzmann constant, the second term is the 
dissipative energy rate (\ref{disener}). In steady case, disk temperature 
is determined

\be
\label{stefbol}
\sigma T^4={3\over 2}{{\cal R}\over \mu}T\Psi\Bigl[
\12(\gamma-1){\rho\langle u^2\rangle\over p}-\gamma\Bigr]+\12 \nu\Sigma
(R\Omega')^2.
\ee 
The mean square velocity is $\langle u^2\rangle=u_\phi^2+u_c^2$, where 
$u_c$ is the velocity dispersion of clupms encountering the disk, for 
clumps orbiting around the central star or falling from external molecular 
cloud $u_c\sim u_\phi$, so we will assume $\langle u^2\rangle \simeq 
2u_\phi^2$; the contribution from radial disk velocity $u_R$ is neglected.  
The corresponding contribution to disk luminosity from the first term in 
the r.h.s. of (\ref{clumheat}) is 

\be
L_D^c=6\pi{{\cal R}\over \mu}\int\limits_{R_\ast}^\infty 
T\Psi\Bigl[\12 (\gamma-1){\rho \langle u^2\rangle \over p}-\gamma\Bigr]
RdR, 
\ee
and when compared to the viscous luminosity in the absence of mass loading
it is 

\be
{L_D^c\over L_{D0}^\nu}\sim {R^2 \Psi\over \nu \Sigma}\sim 
{\dot M_p\over \dot M_0}. 
\ee

In dimensionless units Eq. (\ref{stefbol}) reads as 

\be
\label{temperature}
\theta^4+6\delta{\Lambda-1\over I} \psi\theta=
2(\gamma-1){\Lambda-1\over I}{\psi\over x}+{\Lambda\over x^3}
\left(1-{1\over \sqrt{x}}\right)-
{\Lambda-1\over I}{1\over x^3}\int\limits_1^x \psi xdx, 
\ee
where $x=R/R_\ast$, $\Psi$ assumed in the form $\Psi=\Psi_0\psi(x)$, 
$\theta=T/T_\ast$, $\delta=(c_\ast/c_M)^2$

\ba
T_\ast=\left({3GM\dot M_0\over 8\pi\sigma R_\ast^3}\right)^{1/4},
\quad
\Lambda={\dot M_\ast\over \dot M_0}, \quad \dot M_\ast=2\pi \Psi_0R_\ast^2I+
\dot M_0,
\nonumber\\ 
\quad I=\int\limits_1^\infty \psi xdx,\quad
c_\ast^2=\gamma{{\cal R}\over \mu}T_\ast,\quad 
c_M^2={GM\over R_\ast}.
\ea
The problem contains two dimensionless parameters, $\delta$, and $\Lambda$. 
It is readily seen that $\delta\ll 1$: for $R_\ast=R_\odot$, $M=\msun$ and 
$\dot M_0\sim 10^{-6}~\msun$ yr$^{-1}$ this ratio is $\delta\simeq 10^{-3}$. 
For weak loading, \ie $\dot M_\ast\simgt \dot M_0$, $\Lambda-1$ is also 
small, however even in this case mass loading can change significantly 
temperature distribution in the disk and its spectrum. In Fig. 2 we 
show the temperature distribution $\theta(x)$ for the mass-loading function 

\be
\label{massfunc}
\psi(R)=\Bigl[{R-R_\ast\over R_L}\Bigr]\exp\left(-{R-R_\ast\over R_L}+1\right),
\ee
with the peak $\psi(R)=1$ at $R=R_\ast+R_L$, where $R_L$ is the scale of 
the radial distribution of matter infall, 
$R_L=LR_\ast$, $L=30$ and a set of $\Lambda$. 

It is seen that the most important contribution from additional mass input 
comes from regions $x\simgt L$ where temperature can increase by factor of 
2 even for a moderate mass loading $\Lambda=2$. At larger distances, $x\sim 
10-20~L$, the temperature profiles for $\Lambda>1$ approach the standard 
profile ($\Lambda=1$), and then go lower because at such 
distances the heat input from kinetic energy of clumps ($u^2\propto x^{-1}$, 
first term in the r.h.s. of Eq. \ref{temperature}) is smaller than the energy 
required to heat the infalling (presumably cold) gas (second term in the 
l.h.s. of 
Eq. \ref{temperature}). Similar changes are seen for more extended in radius 
mass input, $L=100$ (Fig. 3). In this case, the surface mass input rate 
for equal $\Lambda$ is $\sim 3$ times lower than for $L=30$, as $\Lambda
-1\propto \Psi_0L$, however the increase in total luminosity of loaded disks 
from the area $\propto L^2$ must be larger. For mass input in the form 
(\ref{massfunc}) and $\Lambda-1\sim 1$ at 
$x<L/3$ the second term in the r.h.s. of (\ref{temperature}) dominates and 
the first and third terms contribute only around 0.01 \% to 1\%, at 
$x\simgt L/3$ the situation changes and main contribution (up to 97\%) 
is from the first tirm in (\ref{temperature}). At $x>10 L$ the dominant 
contribution is again from the second term; the third term at $x>L/3$ remains 
always about 30-50 \% of the second. For $\Lambda-1\sim 0.1$ the second term 
dominates up to $x\sim 2L/3$ and then at $x>8 L$, the last term always 
remains as small as 5-10 \% of the second term. 

In order to illustrate how the mass loading can change the spectra of 
accretion disks, we show here the spectral energy distribution (SED) for 
optically thick disks seen face-on

\be
\label{spectrum}
\Sigma_\nu=2\pi\int\limits_{R_\ast}^{R_{out}}B_\nu[T(R)]RdR,
\ee
with $T(R)=T_\ast \theta(x)$, $B_\nu[T]$ being the Planck function, 
$R_{out}$, the outer radius of the disk, 
in the models shown in Figs (2)--(5) $R_{out}=500$, contribution from 
regions $R>500$ is small. Defining dimensionless frequency $y$ as

\be
\label{frequency} 
y={h\nu\over k_BT_\ast},
\ee
one can reduce (\ref{spectrum}) to the form

\be
\label{spectrumdim}
\Sigma_\nu=\Sigma_0 S(y),
\ee
where 

\be
\Sigma_0={16\pi^2k_B^3T_\ast^3R_\ast^2\over h^2c^2}.
\ee
Fig. 4 shows the spectra from disks with temperature distribution shown in 
Fig. 2. It is seen that even for relatively small loading, $\Lambda=2$, 
the spectrum changes substantially: in maximum it increases by factor of 
5, the maximum itself is more sharp showing an excess of 
energy in this range (for $R=R_\odot$, $M=\msun$ and $\dot M_0=10^{-6}$ 
$\msun$ yr$^{-1}$ the frequency at maximum is $\nu\sim 1-1.5\times 10^{13}$ 
Hz, or $\lambda\sim 10-20~\mu$m). For a more extended disk ($L=100$) 
the changes in 
their spectra are more pronounced as shown in Fig. 5: at $\Lambda=2$ in 
maximum it reaches an order of magnitude increase in comparison with 
mass-unloaded disks, the maximum itself shifts to a lower frequencies, 
and the energy excess in this frequency range is more obvious. These two 
examples show that additional heat input due to mass loading can be 
responsible (or may contribute) for FIR excess observed from circumstellar 
disks around young stars (see for a review Natta, Meyer, Beckwith 1997). In 
order to illustrate such a possibility we compare in Fig. 6 the spectrum 
expected from an optically thick mass-loaded disk seen face-on with the one 
observed in the Chameleon cloud (Natta \etal 1997). For the modelled spectrum 
we assumed the distribution of infalling material 
of $L=9.2$ AU in radius (integration was over $R=3L$), its accretion mass rate 
$\dot M_0=2.2\times 10^{-7}~\msun$ yr$^{-1}$, and the mass loading only 
$\Lambda-1=0.08$, the central star is of $M=1~\msun$, $R_\ast=R_\odot$. Filled 
squares show the FIR excess observed from CS Cha and which cannot be attributed 
to a standard reprocessing accretion disk; open diamonds in optics show 
presumably the photospheric emission. [The lack, or negligibly low level, of 
emission in optical and NIR range from the accretion disk is considered as an 
indication of a hole in the inner disk (about 0.3 AU in size).] 

For a rough estimate of the effects of mass loading on SEDs  
from accretion disks, one can use the following approximate expression 
for temperature in the region where heating due to dissipation of kinetic 
energy of falling clumps (first term in r.h.s. of Eq. \ref{temperature}) 
reaches the maximum (at $x_m\sim \sqrt{L}$ with $(s/x)_m\sim e/L$)

\be
\theta_m\sim \Bigl[2(\gamma-1){\Lambda-1\over I}{e\over L}\Bigr]^{1/4},
\ee
which correctly by order of magnitude describes the asymptotics 
at distances $x$ where accretion 
heating (last two terms in Eq. \ref{temperature}) and cooling due to mixing 
of cold infalling gas (second term in the l.h.s. in Eq. \ref{temperature}) 
are unimportant -- this region corresponds to the plateaus in $\theta(x)$ 
profiles in Figs. 2 and 3. In dimensional units $T_m\sim 0.8
(\Lambda-1)^{1/4}L^{-1/2}T_\ast$. For the above model it gives 
$T_m\sim 100$ K, and the corresponding frequency $\nu_m\sim 6\times 
10^{12}$ Hz.  

\subsection{Effects of angular momentum input}

When infalling gas has non-zero angular momentum equation, so that 
$v_{c,\phi}\neq 0$ equation (\ref{vazimut}) is written in the form 
(Safronov \& Vityazev 1983, Papaloizou \& Lin 1995)

\be
\label{vazload}
{d\over dR}(\Sigma R^3\Omega u_R)={d\over dR}\left(
\nu\Sigma R^3{d\over dR}\Omega\right)+\Psi(R)R^2v_{c,\phi}(R),
\ee
and the contribution to energy equation is determined by 

\be
{2\pi\over \dot M_0R^2\Omega}\int\limits_{R_\ast}^{R}
\Psi(R)R^2v_{c,\phi}(R)dR.
\ee
In an extreme case when the angular velocity of the infalling gas approaches 
the keplerian value $v_{c,\phi}(R)=\sqrt{GM/R}$ energy equation reads as

\ba
\label{tempmomen}
\theta^4+6\delta{\Lambda-1\over I} \psi\theta=
2(\gamma-1){\Lambda-1\over I}{\psi\over x}+{\Lambda\over x^3}
\left(1-{1\over \sqrt{x}}\right)
\nonumber\\
- {\Lambda-1\over I}{1\over x^3}\int\limits_1^x \psi xdx+
{\Lambda-1\over I}{1\over x^{7/2}}\int\limits_1^x\psi x^{3/2}dx. 
\ea
It is readily seen that for the mass input rate in the form (\ref{massfunc}) 
the contribution from the last term is of the same 
order of magnitude (about 2/3) as of the third term, and thus never exceeds 
30\% for $\Lambda\sim 1$ and 3--6\% for $\Lambda-1\ll 1$.

\section{Conclusions }

Dissipation of kinetic energy of clumpy gas falling onto accretion disks 
from the debries of a parent protostellar molecular cloud significantly 
alters the radial temperature distribution. It can be important in outer parts 
of a reprocessing disk, where temperature decreases sufficiently 
$T\sim R^{-3}$, while the heat input rate associated with mass loading 
is approximately proportional to the square of the azimuthal velocity 
$\dot E\propto R^{-1}$. Thus, mass loading can naturally produce excess 
of emission in the long-wavelength range of SEDs of the accretion disks, 
and with an appropriate 
radial distribution of mass loading can reproduce observed FIR excess. 
For a one-peak radial distribution of the infall rate localized at $R\sim L$, 
the corresponding temperature in the hot ring for $R_\ast=1~R_\odot$, 
$M_\ast=\msun$ is 

\be
T_m\sim 800\left({\dot M_p\over \dot M_0}\right)^{1/4}
\left({L\over 1~{\rm AU}}\right)^{-1/2}\left({\dot M_0\over 
10^{-6}~\msun~{\rm yr}^{-1}}\right)^{1/4}~{\rm K},
\ee
and the wavelength 

\be
\label{wavelen}
\lambda_m\sim 6\left({L\over 1~{\rm AU}}\right)^{1/2}
\left({\dot M_p\over \dot M_0}\right)^{-1/4}
\left({\dot M_0\over 10^{-6}~\msun~{\rm yr}^{-1}}\right)^{-1/4}
~\mu{\rm m}.
\ee
It is readily seen that the extra luminosity produced by the infalling 
gas is proportional to the mass loading rate, $L(\dot M_p)\sim \sigma T_m^4
L^2\propto \dot M_p$, and for weak mass loading, $\dot M_p\ll\dot M_0$, is 
small as compared to the total luminosity of an unloaded disk. However, it 
can result in an observable effect if the loading mass is deposited at 
large radii, so that the wavelength in the peak of emission (\ref{wavelen}) 
falls into FIR range where the reprocessing disk does not contribute 
singnificantly. This case is shown in Fig. 6 where for mass loading 
$\dot M_p\sim 0.1 \dot M_0$ the FIR luminosity of the infalling gas is 
approximately ten times smaller than the optical luminosity 
of the reprocessing disk.     

The results presented in this paper have only an illustrative 
meaning, and can be directly applied for explanation of spectral features 
observed in protoplanetary disks only when clear evidences of mass loading 
will be found. Nonetheless, they definitely show that mass loading can have 
an important influence on temperature distribution in accretion disks and 
their spectra. 
\bigskip

I thank S. A. Lamzin, A. V. Lapinov, A. M. Sobolev, I. I. Zinchenko for 
valuable discussions. This work was supported by RFBR (project No 99-02-16938) 
and INTAS (project No 1667). This research has made use of NASA's Astrophysics 
Data System Abstract Service.


\vskip 1cm
{\it Note added in manuscript 2002 August 8.--} After acceptance of this 
paper I have been informed about the papers by Falcke H. \& Melia F., 
1997, {\it Astrophys. J.,} 479, 740, and Coker R. F., Melia F., \& Falcke H., 
1999, {\it Astrophys. J.,} 523, 642 which deal with accretion disks around 
massive black holes loaded by wind infall with application to Sgr A$^*$,  
in the framework of hydrodynamical equations similar to that used in this 
paper. Their emission spectrum has an excess in the long-wavelength limit 
due to additional energy input connected with the mass-loading similar to 
that obtained in our paper. 

\newpage

\begin{figure}
\epsfxsize=6truein
\epsfysize=4truein
\epsfbox{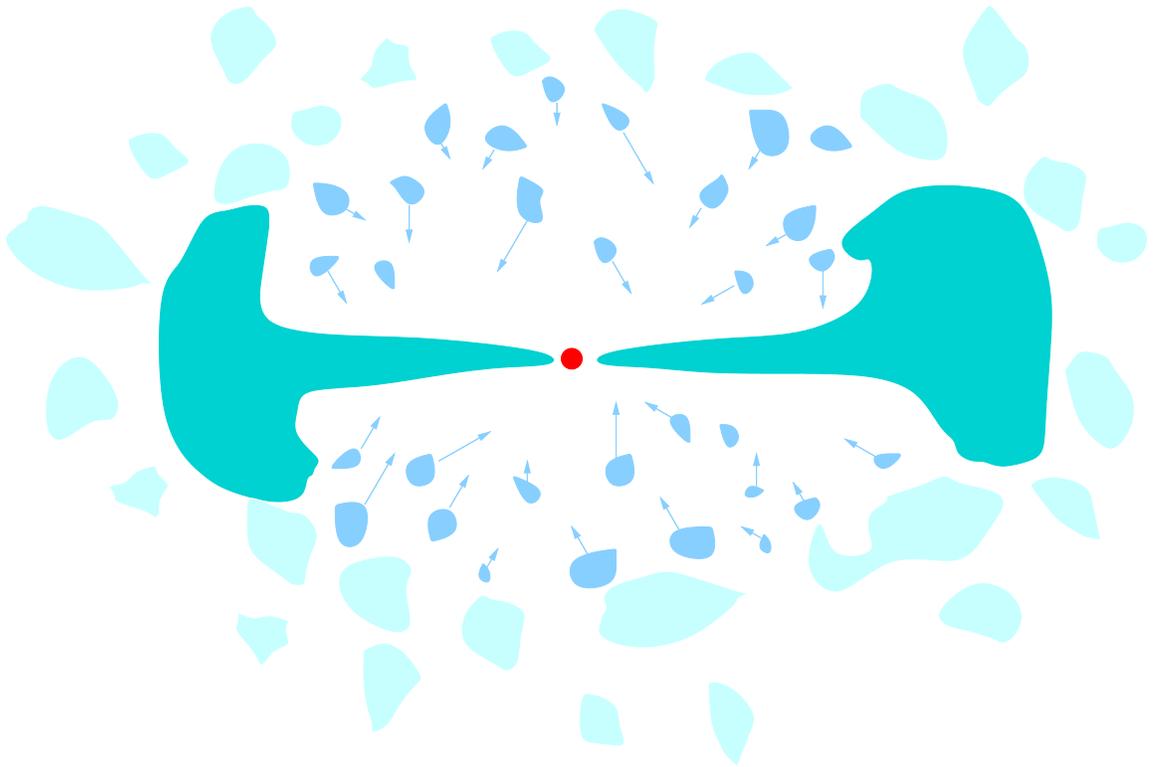}
\caption{Schematic representation of the model: debries of the seed cloud 
falling down on the accretion protoplanetary disk.}
\end{figure}

\begin{figure}
\epsfxsize=5truein
\epsfysize=5truein
\epsfbox{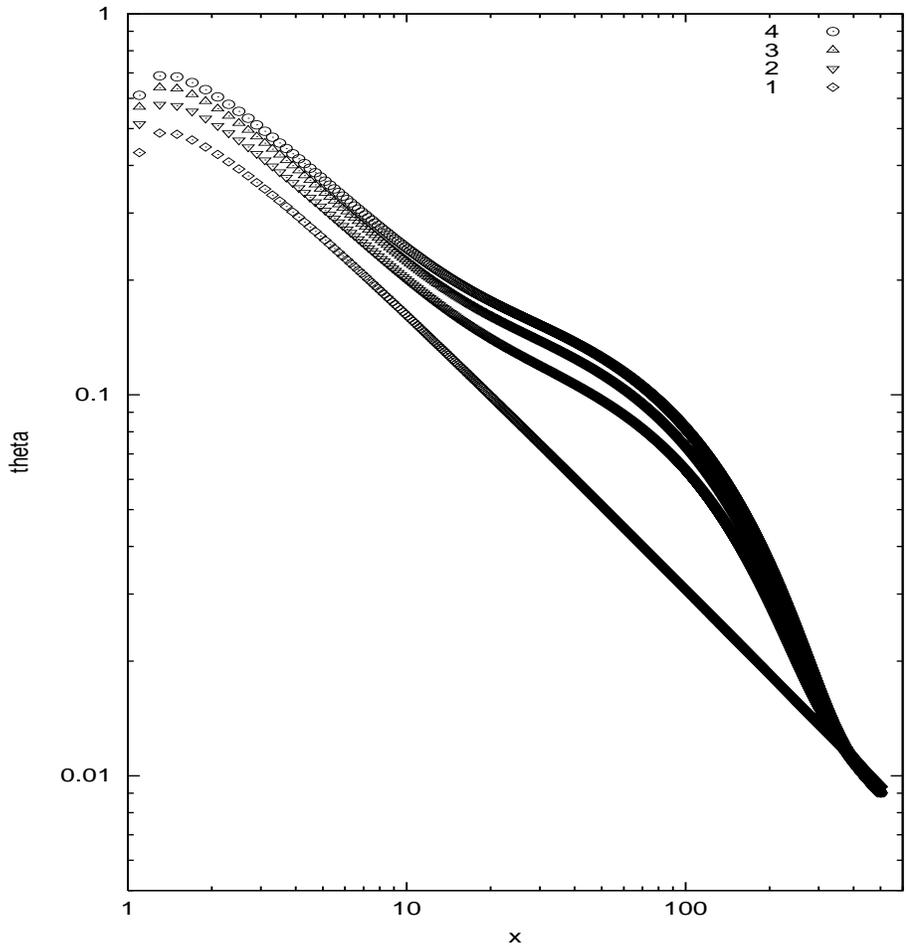}
\caption{Temperature distribution $\theta(x)$ for the mass-loading function 
(\ref{massfunc}), $\delta=10^{-3}$, $\Lambda=1$ (unloaded), 2, 3, 4 from 
the bottom to top, $L=30$.}
\end{figure}

\begin{figure}
\epsfxsize=5truein
\epsfysize=5truein
\epsfbox{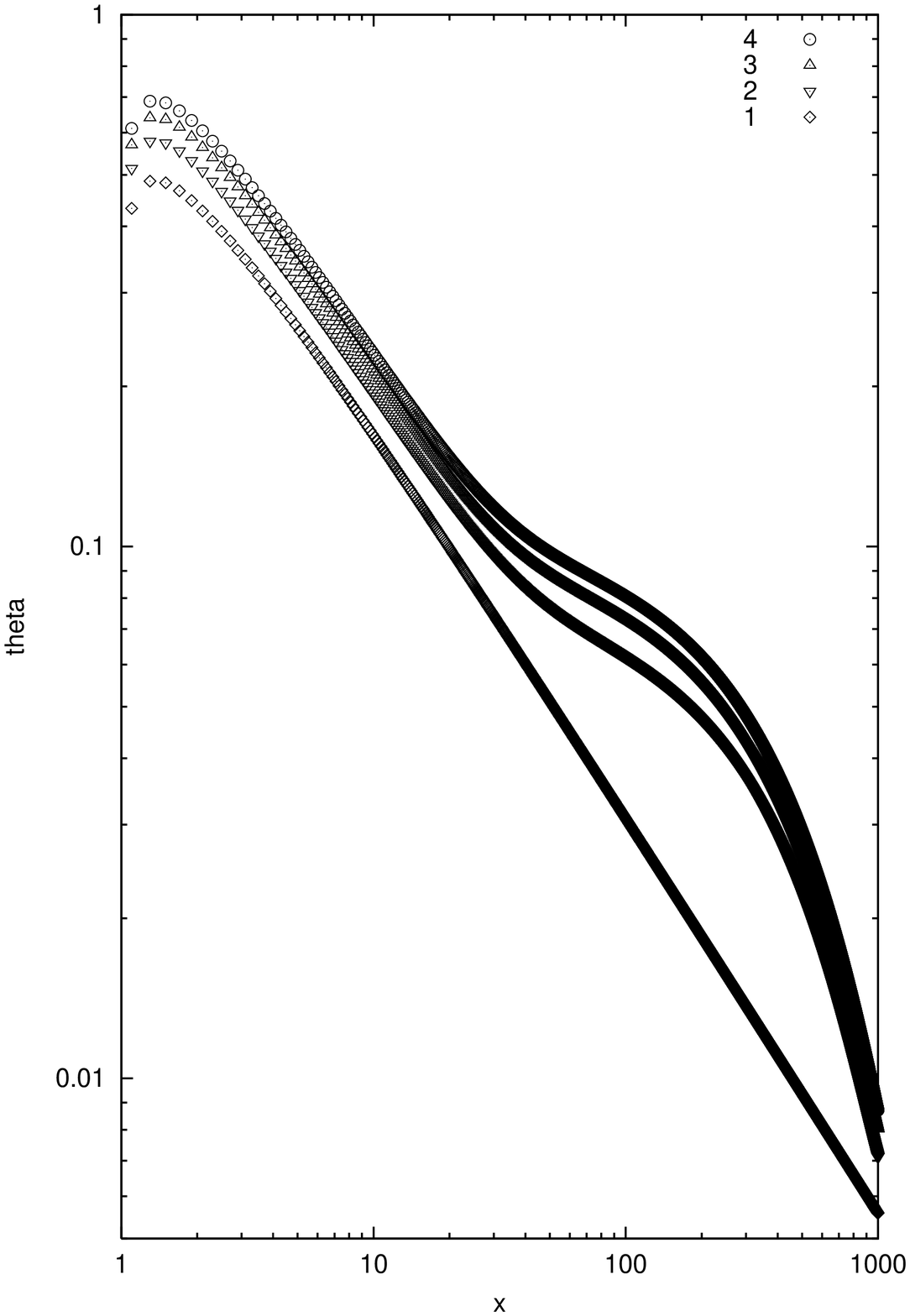}
\caption{Same as in Fig. 2 for $L=100$.}
\end{figure}

\begin{figure}
\epsfxsize=5truein
\epsfysize=5truein
\epsfbox{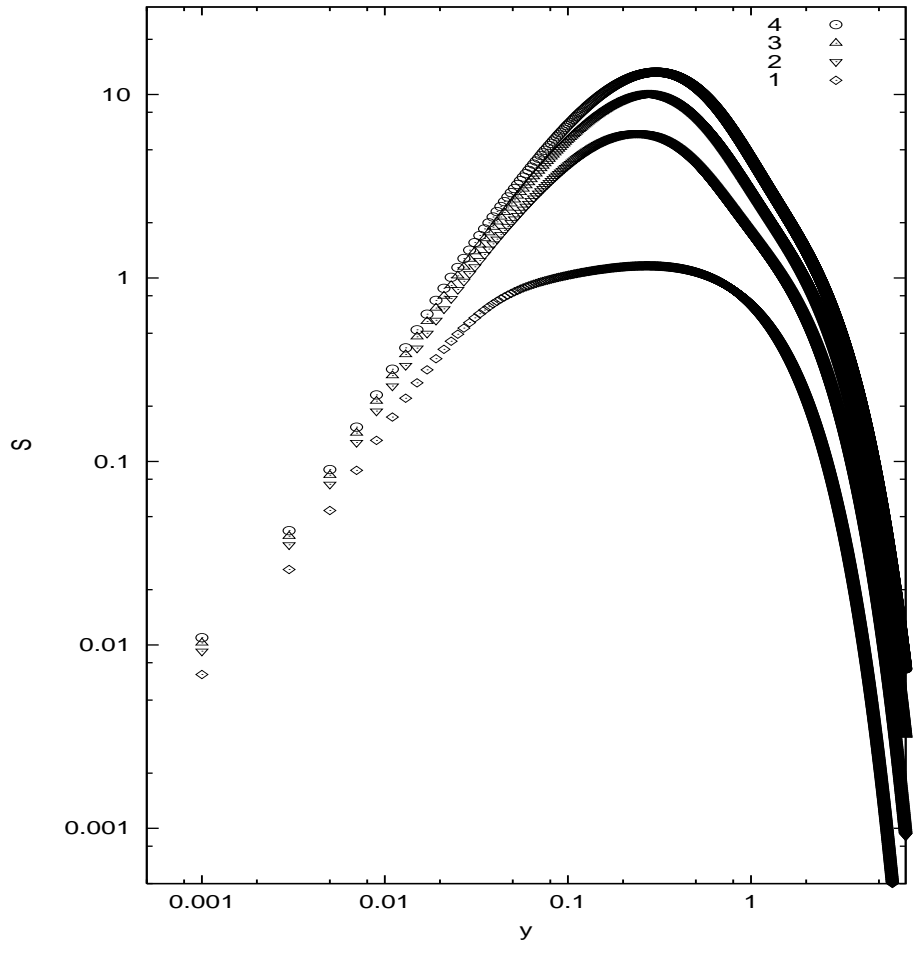}
\caption{Energy spectrum for $\theta(x)$ shown in Fig. 2.}
\end{figure}

\begin{figure}
\epsfxsize=5truein
\epsfysize=5truein
\epsfbox{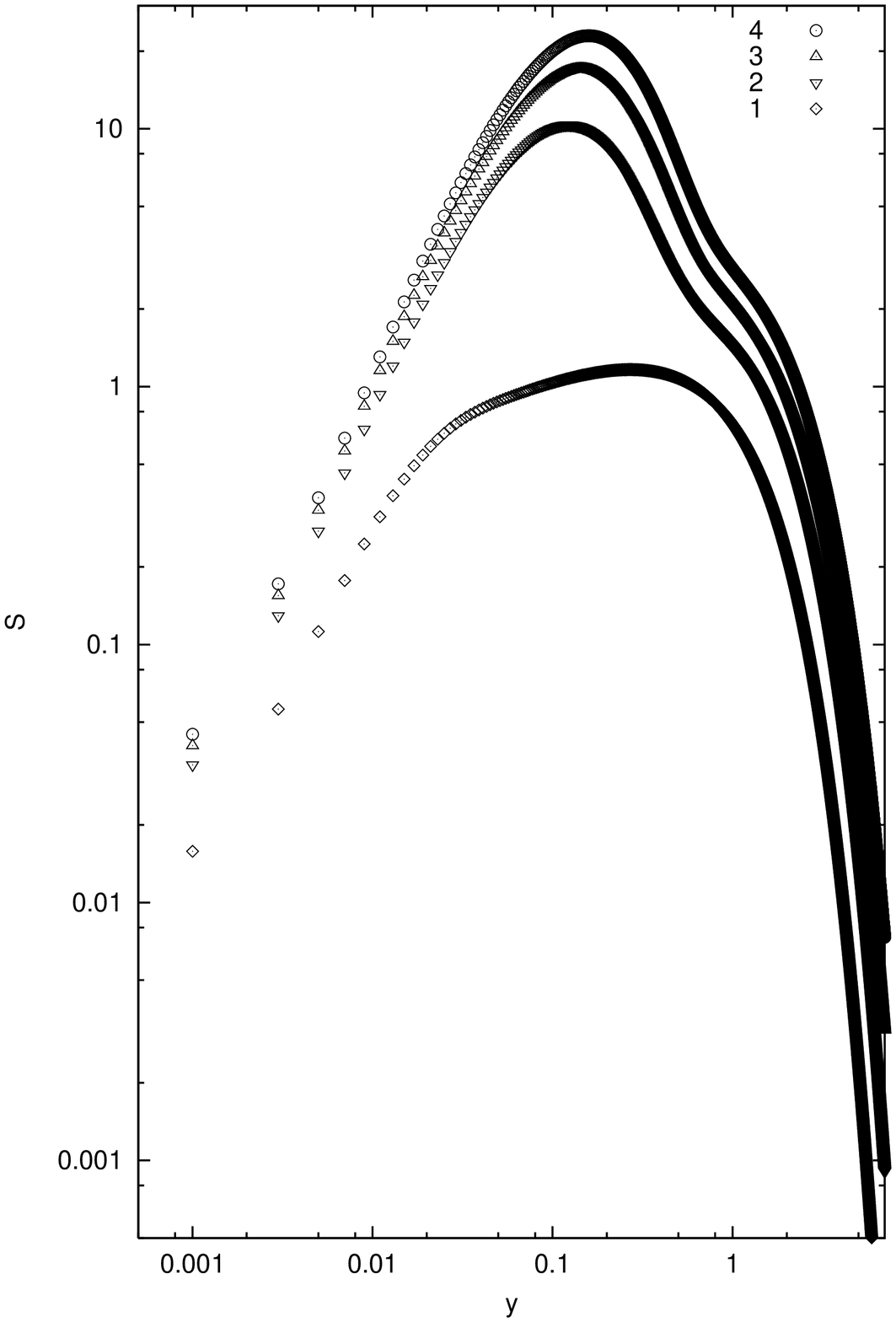}
\caption{Same as in Fig. 4 for $L=100$.}
\end{figure}

\begin{figure}
\epsfxsize=5truein
\epsfysize=5truein
\epsfbox{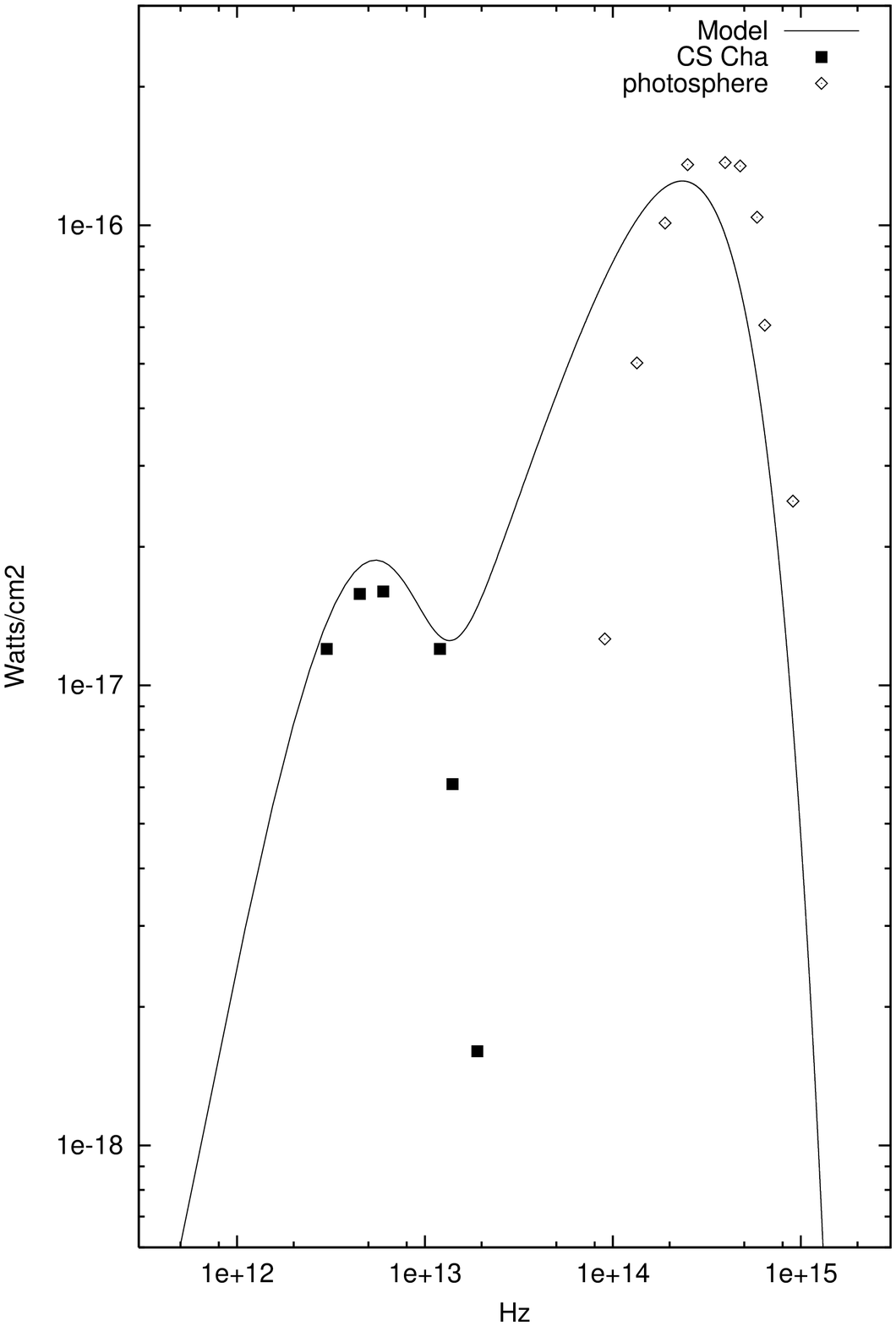}
\caption{SED in CS Cha: open diamonds from 
ground-based observations (presumably represent the spectrum of the 
stellar photosphere), filled squares from ISO and IRAS -- Natta, Meyer, \& 
Beckwith (1997), solid line shows the spectrum of a mass-loaded disk 
with $\Lambda=1.08$, $L=9.2$ AU, $\dot M_0=2.2\times 10^{-7}~\msun$ yr$^{-1}$.}
\end{figure}


\begin{thebibliography}{}

\bibitem{} Barnes, S., Tobin, W, Pollard, K. R. 2000, The variable CaII absorption in 
beta-Pictoris during 1998, Publ. Astron. Soc. Australia, 17, 241

\bibitem{} Cowie, L. L., McKee, C. F., Ostriker, J. P. 1981, Supernova remnant revolution 
in an inhomogeneous medium. I - Numerical models, Astrophys. J., 247, 908

\bibitem{} Falgarone, E., Phillips, T. G., Walker, C. K. 1991, The edges of molecular 
clouds - Fractal boundaries and density structure, Astrophys. J., 378, 186

\bibitem{} Grinin, V., Natta, A., Tambovtseva, L. 1996, Evaporation of star-grazing 
bodies in the vicinity of UX Ori-type stars, Astron. \& Astrophys., 313, 857 

\bibitem{} Hartquist, T. W., Dyson, J. E., Pettini, M., Smith, L. J. 1986, 
Mass-loaded astronomical flows. I - General principles and their application 
to RCW 58, Month. Not. Roy. Astron. Soc., 221, 715

\bibitem{} Klein, R. I., McKee, C. F., Colella, P. 1994, On the hydrodynamic interaction 
of shock waves with interstellar clouds. 1: Nonradiative shocks in small 
clouds, Astrophys. J., 420, 213

\bibitem{} Lecavelier des Etangs, A., Hobbs, L. M., Vidal-Madjar, A., Beust, 
H., Feldman, P. D., Ferlet, R., Lagrange, A.-M., Moos, W., \& McGrath, M. 2000, 
Possible emission lines from the gaseous $\beta$ Pictoris disk, Astron. \& Astrophys., 356, 691

\bibitem{} Lynden-Bell, D., Pringle, J. E. 1974, The evolution of viscous discs and the 
origin of the nebular variables, Month. Not. Roy. Astron. Soc., 168, 603

\bibitem{} Natta, A., Meyer, M. R., \& Beckwith, S. V. W. 1997, Circumstellar 
disks around pre-main-sequence stars: what ISO can tells us, In: {\it Star 
Formation with the Infrared space Observatory}, J. Zun \& R. Liseau, eds, 
ASP Conference Series, v. 132, p. 265

\bibitem{} Papaloizou, J. C. B., Lin, D. N. C. 1995, Theory Of Accretion Disks I: Angular 
Momentum Transport Processes, Ann. Rev. Astron. Astrophys., 33, 505

\bibitem{} Pringle, J. E. 1981, Accretion discs in astrophysics, 
Ann. Rev. Astron. Astrophys., 19, 137

\bibitem{} Roberge, A., Feldman, P. D., Lagrange, A. M., Vidal-Madjar, A., 
Ferlet, R., Jolly, A., Lemaire, J. L., Rostas, F. 2000, High-Resolution Hubble 
Space Telescope STIS Spectra of C I and CO in the $\beta$ Pictoris 
Circumstellar Disk, Astrophys. J., 538, 904

\bibitem{} Safronov, V. S., Vityazev, A. V. 1983, The origin of Solar 
system, In: {\it  Astrophysics and Cosmic Physics}, R. A. Sunyaev, ed., 
p. 5 

\bibitem{} Shakura, N. I., Sunyaev, R. A. 1973, Black holes in binary systems. 
Observational appearance, Astron. Astrophys., 24, 337

\end{thebibliography}
\end{document}